\documentclass{ws-mpla}

\begin{document}

\markboth{U. Mukhopadhyay, P. C. Ray, S. Ray and S. B. Dutta
Choudhury} {Generalization of Phenomenological $\Lambda$ Model for
Dark Energy Investigation}

\catchline{}{}{}{}{}
%
%%%%%%%%%%%%%%%%%%%%%%%%%%%%%%%%%%%%%%%%%%%%%%%%%%%%%%%%%%%%%%%%%%%%

\title{Generalized Model for $\Lambda$-Dark Energy}

\author{UTPAL MUKHOPADHYAY}

\address{Satyabharati Vidyapith, Barasat, North 24 Parganas,
Kolkata 700 126, West Bengal, India}

\author{P. C. RAY}

\address{Department of Mathematics, Government College of
Engineering and Leather Technology, Kolkata 700 098, West Bengal,
India; raypratap1@yahoo.co.in}

\author{SAIBAL RAY}

\address{Department of Physics, Barasat Government College, Kolkata 700 124,
North 24 Parganas, West Bengal, India; saibal@iucaa.ernet.in}

\author{S. B. DUTTA CHOUDHURY}

\address{Department of Physics, Jadavpur University,
  Kolkata 700 032, West Bengal, India}

\maketitle

\pub{Received (Day Month Year)}{Revised (Day Month Year)}

\begin{abstract}
Einstein field equations under spherically symmetric space-times
are considered here in connection to dark energy investigation. A
set of solutions are obtained for a kinematical $\Lambda$ model,
viz., $\Lambda \sim (\dot a/a)^2$ without assuming any {\it a
priori} value for the curvature constant and the equation of state
parameter $\omega$. Some interesting results, such as the nature
of cosmic density $\Omega$ and deceleration parameter $q$, have
been obtained with the consideration of two-fluid structure
instead of usual uni-fluid cosmological model.

\end{abstract}

\ccode{PACS Nos.: 98.80.Jk, 98.80.Cq.}

\section{Introduction}

Present-day cosmological research hovers around the investigation
of {\it dark energy}, an exotic type of entity responsible for
generating acceleration in expanding Universe. In fact, various
recent observational results
\cite{Riess1998,Perlmutter1999,Knop2003,Spergel2003,Riess2004,Tegmark2004,Astier2005,Spergel2006}
suggest that the Universe is expanding with an acceleration alone
while some other works
\cite{Riess2001,Amendola2003,Padmanabhan2003} indicate that the
acceleration is a phenomenon of recent past and was preceded by a
phase of deceleration.

Now, the exact nature of dark energy being still unknown, its
investigation are going on along various paths. Phenomenological
models are also contenders in this dark energy investigation.
Although these type of phenomenological models do not originate
from any underlying quantum field theory, yet they are useful
enough to arrive at some fruitful conclusions. Out of three main
variants of phenomenological models, viz. kinematic, hydrodynamic
and field-theoretic models \cite{Sahni2000}, the present work
deals with kinematical models where the dark energy representative
$\Lambda$ is assumed to be a function of time. Recently, Ray et
al.\cite{Ray2007} and Mukhopadhyay et al.\cite{Mukhopadhyay2005}
have shown the equivalence of four $\Lambda$ models, viz. $\Lambda
\sim (\dot a/a)^2$, $\Lambda \sim \ddot a/a$, $\Lambda \sim \rho$
and $\Lambda\sim \dot H$ for spatially flat ($k=0$) Universe. But,
since the closed ($k=1$) and open ($k=-1$) Universes cannot be
entirely ruled out, so there is enough reason to investigate dark
energy for general $k$. In this work, therefore, one of the
equivalent $\Lambda$ models, viz. $\Lambda \sim (\dot a/a)^2$ is
selected to solve Einstein equations for general $k$ in order to
have a broader view of accelerating Universe.

The scheme of the investigation is as follows: Sec. 2 and 3 deals
respectively with the Field equations and their solutions while
some physical feature arising out of this work are described in
Sec. 4. Finally, some conclusions are made in Sec. 5.

\section{Field Equations for the Spherically Symmetric Space-times}
The Einstein field equations are given by
\begin{eqnarray}
R^{ij}-\frac{1}{2}Rg^{ij}= -8\pi G\left[T^{ij}-\frac{\Lambda}{8\pi
G}g^{ij}\right]
\end{eqnarray}
where the cosmological term $\Lambda$ is time-dependent, i.e.
$\Lambda = \Lambda(t)$.

Let us choose the spherically symmetric FLRW metric
\begin{eqnarray}
ds^2=
-dt^2+a(t)^2\left[\frac{dr^2}{1-kr^2}+r^2(d\theta^2+sin^2\theta
 d\phi^2)\right]
\end{eqnarray}
where $k$ is the curvature constant and  $a=a(t)$ is the scale
factor.
 For the metric given by equation (2), the field equations (1)
 yield  Friedmann and Raychaudhuri equations respectively given by
\begin{eqnarray}
\left(\frac{\dot a}{a}\right)^2+\frac{k}{a^2} = \frac{8\pi
G\rho}{3}+\frac{\Lambda}{3},
\end{eqnarray}
 \begin{eqnarray}
\left(\frac{\ddot a}{a}\right) = -\frac{4\pi
G}{3}\left(\rho+3p\right)+\frac{\Lambda}{3}
\end{eqnarray}
where $c$, the velocity of light in vacuum, is assumed to be
unity.

The generalized  energy conservation law, when both $\Lambda$ and
$G$ vary, is derived by Shapiro et al.\cite{Shapiro2005} using
Renormalization Group Theory as well as by Vereschagin and
Yegorian\cite{Vereshchagin2006} using Gurzadyan-Xue
formula\cite{Gurzadyan2003}. Since in the present work $G$ is
assumed as a constant and $\Lambda$ is a variable, then the above
mentioned generalized conservation law reduces to the particular
form
\begin{eqnarray}
8\pi G(p+\rho)\left(\frac{\dot a}{a}\right) = -\frac{8\pi
G}{3}\dot\rho-\frac{\dot\Lambda}{3}.
\end{eqnarray}
The barotropic equation of state relating pressure and density is
given by
\begin{eqnarray}
p= \omega\rho
\end{eqnarray}
where the barotropic index $\omega$ can assume the values $0$,
$1/3$, $1$ and $-1$ for pressure-less dust, electromagnetic
radiation, stiff fluid and vacuum fluid respectively.

From (1), using equation (6), we get,
\begin{eqnarray}
\rho =\frac{3}{4\pi
G(1+3\omega)}\left(\frac{\Lambda}{3}-\frac{\ddot a}{3}\right)
\end{eqnarray}
Again, differentiating equation (3) and using equations (5) - (7)
we obtain the differential equation
\begin{eqnarray}
\left(\frac{\dot
a}{a}\right)^2+\frac{2}{1+3\omega}\left(\frac{\ddot
a}{a}\right)+\frac{k}{a^2} =
\left(\frac{1+\omega}{1+3\omega}\right)\Lambda.
\end{eqnarray}

\section{Solutions for the Phenomenological Model $\Lambda = 3\alpha (\dot a/a)^2$}

Using the {\it ansatz} $\Lambda = 3\alpha (\dot a/a)^2$, we
immediately get from equation (8)
\begin{eqnarray}
\frac{\ddot a}{\dot a} =
\left[3\alpha(1+\omega)-(3\omega+1)\right]\frac{\dot
a}{2a}-(3\omega+1)\frac{k}{2a\dot a}.
\end{eqnarray}
The above equation after simplification reduces to the form
\begin{eqnarray}
a\dot a \frac{d}{dt}\left[ln{(\dot a a^{-s/2})}\right] =
-\frac{(3\omega+1)k}{2}
\end{eqnarray}
where $s = 3\alpha(1+\omega)-(3\omega+1)$.

Let us now study the following case when $s = -2$. In this case
equation (10) reduces to
\begin{eqnarray}
a\dot a \frac{d}{dt}\left[ln(a\dot a)\right] =
-\frac{(3\omega+1)k}{2}.
\end{eqnarray}
Solving equation (11) we get our solution set as
\begin{eqnarray}
a(t) =
\left[{C_0}^{\prime}t+{C_1}^{\prime}-\frac{(3\omega+1)}{2}kt^2\right]^{1/2},
\end{eqnarray}
\begin{eqnarray}
H(t) =
\frac{{C_0}^{\prime}-(1+3\omega)kt}{2\left[{C_0}^{\prime}t+{C_1}^{\prime}-\frac{(3\omega+1)}{2}kt^2\right]},
\end{eqnarray}
\begin{eqnarray}
\rho(t) = \frac{3(1-3\alpha)}{16\pi
G}\frac{\left[{C_0}^{\prime}-\frac{(3\omega+1)}{2}kt\right]+
2k\left[{C_0}^{\prime}t+{C_1}^{\prime}-\frac{(3\omega+1)}{2}kt^2\right]}
{\left[[{C_0}^{\prime}t+{C_1}^{\prime}-\frac{(3\omega+1)}{2}kt^2\right]^2},
\end{eqnarray}
\begin{eqnarray}
\Lambda(t) =
\frac{3\alpha[{C_0}^{\prime}-(3\omega+1)kt]^2}{4[{C_0}^{\prime}t+{C_1}^{\prime}-\frac{(3\omega+1)}{2}kt^2]^2}
\end{eqnarray}
where ${C_0}^{\prime}=2C_0$, ${C_1}^{\prime}=2C_1$, $C_0$ and
$C_1$ being constants of integration.

If we impose the boundary condition $a(t)=0$ when $t=0$, then
$C_1=0$ which implies ${C_1}^{\prime}=0$. Then the simplified
solution set becomes
\begin{eqnarray}
a(t) =
\left[{C_0}^{\prime}t-\frac{(3\omega+1)}{2}kt^2\right]^{1/2},
\end{eqnarray}
\begin{eqnarray}
H(t) =
\frac{{C_0}^{\prime}-(1+3\omega)kt}{2\left[{C_0}^{\prime}t-\frac{(3\omega+1)}{2}kt^2\right]},
\end{eqnarray}
\begin{eqnarray}
\rho(t) =
3\frac{[(\alpha+1){{C_0}^{\prime}-(1+3\omega)kt}^2+2(1+3\omega)k{{C_0}^{\prime}t-\frac{(1+3\omega)}{2}kt^2}]}{16\pi
G (1+3\omega)[{C_0}^{\prime}t-\frac{(1+3\omega)}{2}kt^2]^2},
\end{eqnarray}
\begin{eqnarray}
\Lambda(t) =
\frac{3\alpha\left[{C_0}^{\prime}-(3\omega+1)kt\right]^2}{4\left[{C_0}^{\prime}t-\frac{(3\omega+1)}{2}kt^2\right]^2}.
\end{eqnarray}
It is clear from equation (19) that for a repulsive $\Lambda$,
$\alpha$ must be positive whereas $\alpha=0$ implies a null
$\Lambda$. This means that we are getting Einstein's expanding
Universe without $\Lambda$.

\section{Physical Features of the Solutions}
\subsection{Density of the Universe $\Omega$} The above solution
set is obtained by assuming $s=-2$. Now, $s=-2$ means
\begin{eqnarray}
\frac{2}{3(1-\alpha)(1+\omega)} = \frac{1}{2}.
\end{eqnarray}
For $k=0$ we get from equations (16), (17), (19) and (20)
respectively $a(t)\propto t^{2/3(1-\alpha)(1+\omega)}$,
$H(t)\propto 1/t$, $\rho(t)\propto 1/t^2$ and $\Lambda(t)\propto
1/t^2$. These results were obtained by Ray et al.\cite{Ray2007}
for flat ($k=0$) Universe. Again, from equation (20) we have
\begin{eqnarray}
\frac{(3\omega-1)}{(1+\omega)} = 3\alpha.
\end{eqnarray}

Since equation (19) suggests that for a repulsive $\Lambda$ we
must have $\alpha>0$, then from equation (21) we find that either
$\omega>1/3$ or $\omega<-1$. For $\omega>1/3$ we get a Universe
where contribution of electromagnetic radiation is negligible (for
radiation dominated Universe, $\omega=1/3$) while $\omega<-1$
signifies the presence of phantom energy. Again, using equation
(18), the expression for cosmic matter energy density $\Omega_m$
can be easily derived and is given by
\begin{eqnarray}
\Omega_m =
\frac{2(\alpha+1)}{(1+3\omega)}+4k\frac{[{C_0}^{\prime}t-\frac{(1+3\omega)}{2}kt^2]}{[{C_0}^{\prime}-(1+3\omega)kt]^2}.
\end{eqnarray}

Also, from the {\it ansatz} $\Lambda = 3\alpha (\dot a/a)^2$ we
get
\begin{eqnarray}
\Omega_{\Lambda} = \alpha.
\end{eqnarray}

Then, using equations (21) - (23) we obtain
\begin{eqnarray}
\Omega_m + \Omega_{\Lambda} =
1+\frac{4k[{C_0}^{\prime}t-\frac{(1+3\omega)}{2}kt^2]}{[{C_0}^{\prime}-(1+3\omega)kt]^2}.
\end{eqnarray}

For the flat Universe ($k=0$), equation (24) reduces to the case
of Ray et al.\cite{Ray2007}. Again, equation (24) shows that at
time $t=0$, the sum of $\Omega_m$ and $\Omega_{\Lambda}$ becomes
independent of the curvature constant $k$ and takes a unit value
whatever may be the value of $k$. On the other hand, when $t$
tends to infinity, from (24) we have
\begin{eqnarray}
\Omega_m + \Omega_{\Lambda} = 1-\frac{2}{1+3\omega}.
\end{eqnarray}
 From equation (25) we again observe that, $\Omega_m + \Omega_{\Lambda}$ is
 independent of $k$. Thus, both the early and late phases of the
 Universe exhibit the same behaviour so far as the curvature
 dependency of  the sum of $\Omega_m$ and $\Omega_{\Lambda}$ is
 concerned. It has already been shown that for physically valid
 $\alpha$, either $\omega>1/3$ or $\omega<-1$. In the former case,
 $2/(1+3\omega)<1$ and hence by equation (25)
\begin{eqnarray}
0<\Omega_m + \Omega_{\Lambda}<1.
\end{eqnarray}

But, for $\omega<-1$ we have $-2/(1+3\omega)<1$ and hence equation
(25) provides the following constraint
\begin{eqnarray}
1<\Omega_m + \Omega_{\Lambda}<2.
\end{eqnarray}

The above two relations (26) and (27) suggest that in distant
future not only the sum total of matter and dark energy density
will be independent of curvature of space but also they will be
either less than (for $\omega>1/3$) or greater than (for
$\omega<-1$) unity which misfits with the present status of the
sum of two type of energy densities dominating the present
Universe . This result is very important for visualizing the
cosmic evolution in future.

Now, let us suppose that the Universe is composed of a mixture of
two types of fluids having barotropic indices $\omega_a$ and
$\omega_b$ (say). Then, from equation (25) we get
\begin{eqnarray}
(\Omega_m + \Omega_{\Lambda})_a = 1-\frac{2}{1+3\omega_a},
\end{eqnarray}
\begin{eqnarray}
(\Omega_m + \Omega_{\Lambda})_b = 1-\frac{2}{1+3\omega_b}.
\end{eqnarray}

If $(\Omega_m + \Omega_{\Lambda})_{avg}$ be the average of
$(\Omega_m + \Omega_{\Lambda})_a$ and $(\Omega_m +
\Omega_{\Lambda})_b$ then
\begin{eqnarray}
(\Omega_m + \Omega_{\Lambda})_{avg} =
1-\left[\frac{2+3(\omega_a+\omega_b)}{(1+3\omega_a)(1+3\omega_b)}\right].
\end{eqnarray}
Equation (30) shows that when $\omega_a+\omega_b=-2/3$, then
\begin{eqnarray}
(\Omega_m + \Omega_{\Lambda})_{avg} = 1.
\end{eqnarray}
This means that like the early and present Universe, for late
Universe also the sum of $\Omega_m$ and $\Omega_{\Lambda}$  will
be unity only if the Universe contains a mixture of two types of
fluids rather than a single fluid. Since $\omega_a+\omega_b=-2/3$
and it has already been shown that either $\omega>1/3$ or
$\omega<-1$, then let us suppose, $\omega_a=1/3+\epsilon$ and
$\omega_b=-1-\epsilon$ where $\epsilon>0$. Therefore, for large
value of $t$, when the Universe is filled with a mixture of two
types of fluids (one of them being phantom-fluid) and if the value
of barotropic index of one fluid is $-(1/3+\epsilon)$ and that of
the other is $(-1-\epsilon)$, then the average value of the sum of
$\Omega_m$ and $\Omega_{\Lambda}$ will be equal to one.

\subsection{Deceleration parameter $q$}
Let us now consider the expression for the deceleration parameter
$q$ which is given by
\begin{eqnarray}
q = -\frac{a\ddot a}{\dot a^2} = -\left(1+\frac{\dot
H}{H^2}\right).
\end{eqnarray}
If the Universe is composed of two fluids with equation of state
parameters $\omega_a$ and $\omega_b$, then each of them will have
some effect on the dynamics of the Universe. So for calculating
the value of the deceleration parameter $q$, contributions coming
from each component should be taken into account. If $q_a$ and
$q_b$ be the values of two separate parts of $q$ coming from
fluids having barotropic indices $\omega_a$ and $\omega_b$
respectively, then using equation (17) we get from equation (33)
the following two expressions
\begin{eqnarray}
q_a =
\frac{{{C_0}^{\prime}}^2}{[{C_0}^{\prime}-(2+3\epsilon)kt]^2},
\end{eqnarray}
\begin{eqnarray}
q_b =
\frac{{{C_0}^{\prime}}^2}{[{C_0}^{\prime}+(2+3\epsilon)kt]^2}.
\end{eqnarray}
If $q_{eff}$ be the effective value for $q$, coming after
considering the separate parts $q_a$ and $q_b$, then
\begin{eqnarray}
q_{eff} =
\frac{4{{C_0}^{\prime}}^3kt}{[{{C_0}^{\prime}}^2-(2+3\epsilon)^2k^2t^2]^2}.
\end{eqnarray}
Equation (35) shows that the sign of $q$ depends only on two
quantities, viz., the integration constant ${C_0}^{\prime}(=2
C_0)$ and the curvature constant $k$. If for simplicity we assume
$C_0$ to be positive then we get an accelerating or a decelerating
Universe according as $k<0$ or $k>0$. This result can be
interpreted as follows. The Universe is made of two types of
fluids having equation of state parameters $\omega_a$ and
$\omega_b$, one of which is acting as a prohibitor and another as
a supporter of cosmic acceleration. In the previous matter
dominated phase, $k$ had a small positive value (i.e. $q_a>q_b$)
and as a result the Universe was decelerating. But at a certain
time during cosmic evolution, the second type of fluid (viz.,
phantom fluid) took the upper-hand (i.e. $q_a<q_b$) and
consequently $k$ has become slightly negative. That is why the
present Universe is in a state of acceleration. A very small
positive or negative value of the curvature constant do not
contradict the observational result that the present Universe is
nearly flat. Further, the change of sign of $q$ shows that the
cosmic acceleration is a recent phenomenon.

\section{Conclusions}

The present work, apart from being a generalization of an earlier
one\cite{Ray2007}, has revealed some new and interesting physical
features also. For a flat Universe, all the results of Ray et
al.\cite{Ray2007} can be recovered from the expressions of $a(t)$,
$H(t)$, $\rho(t)$ and $\Lambda(t)$ of the present work. Moreover,
it has been possible to trace the entire cosmic evolution,
starting from the Big-Bang and extending to distant future. The
most significant result is related to the cosmic matter and dark
energy densities for non-flat Universe. It has been shown that for
non-flat Universe, $\Omega_m + \Omega_{\Lambda}=1$ only when the
Universe is composed of two types of fluids, one with $\omega>1/3$
and another with $\omega<-1$. This means that both stiff-fluid
($\omega=1$) and phantom-fluid ($\omega=-1$) can be one of the two
constituents of the Universe when the Universe is not flat. The
evolution of the deceleration parameter $q$ shows that a non-flat
Universe would be decelerating in the past and accelerating at
present. This result is very significant for $\Lambda$-CDM
cosmology.

Another interesting point is the absence of Big-reap even in the
presence of a fluid with $\omega<-1$. Caldwell\cite{Caldwell2002}
and Caldwell et al.\cite{Caldwell2003} demonstrated the occurrence
of a Big-reap in the presence of a fluid with supernegative
equation of state. It may be mentioned here that Gonzalez-Diaz
\cite{Gonzalez-Diaz2003} has shown that by a proper generalization
of the Chaplygin-gas model, a Big-reap can be avoided even in the
presence of phantom energy whereas Abdalla et
al.\cite{Abdalla2004} have arrived at the same result through a
slight modification of GTR. But in the present work, a cosmic
doomsday is shown to be impossible within the normal framework of
GTR. One of the reasons of it may be the presence of another fluid
apart from the phantom fluid. That other fluid may act as an
inhibitor of Big-reap. It may be mentioned that
\u{S}tefan\u{c}i{\'c} \cite{Stefancic2004} has developed a model
in which the dark energy component and the matter component
interact with each other resulting in the appearance of phantom
energy out of non-phantom matter. The present work can be
considered as a counter example of that because here phantom
matter, in the presence of another component as a mediator, can
behave as a non-phantom matter.

Finally, it is to be noted that the present work has demonstrated
that although current observational data points towards a $k=0$
Universe, yet we are not in a position right now to completely
rule out $k=\pm 1$ cosmologies.

\section*{Acknowledgments}
 One of the authors (SR) would like to
express his gratitude to the authorities of IUCAA, Pune, India for
providing him the Associateship Programme under which a part of
this work was carried out.

\end{document}